# Effect of Si and Ga substitutions on the magnetocaloric properties of NiCoMnSb quaternary Heusler alloys

Roshnee Sahoo[1], Ajaya K. Nayak[1], K. G. Suresh[1] and A. K. Nigam[2]

[1]Magnetic Materials Laboratory, Department of Physics, Indian Institute of Technology Bombay, Mumbai- 400076, India

[2]Tata Institute of Fundamental Research, Homi Bhabha Road, Mumbai- 400005, India

*The effect of Si and Ga substitutions on the magnetic and the magnetocaloric properties in Heusler based system $Ni_{46}Co_4Mn_{38}Sb_{12-x}Z_x$ (Z=Si and Ga) has been studied. From the M(T) plots it is found that Si substitution stabilizes the austenite phase, whereas, Ga substitution stabilizes the martensite phase. Strong metamagnetic behaviour is observed in the M(H) isotherms for Si=0.75 and 1, whereas, such a behaviour is absent in the Ga substituted alloys. Associated with magneto-structural transition, large MCE of 58 J/kg K and 70 J/kg K is observed for x=0.75 and 1, respectively in the case of Si. Though the MCE observed in x=0.5 and 1 in the case of Ga is much lower, the MCE peak is found to be quite broad.*

## I. INTRODUCTION

Full Heusler alloys are often found with $X_2YZ$ stoichiometry, where X and Y are usually 3d transition metal atoms and Z is a group IIIA-VA atom such as Sn, In, Ga or Sb etc.[1-3] The noticeable feature in these alloys is the observation of martensitic transition, which can be interpreted as a structural phase transition from high temperature and highly ordered austenite phase to low temperature and less ordered martensite phase. Due to the coupling of the structural transition with the magnetic transition, many interesting properties like magnetocaloric effect[4,5] and shape memory effect[2,6] are observed in the system. Among all Heusler alloys, NiMn based full Heusler systems have been found to be the most exciting. The magnetic properties in this system are mainly due to Mn magnetic moment as Ni magnetic moment is negligible. The Mn-Mn indirect exchange interaction (RKKY type) plays the key role for the magnetism in this system.[7] Recently off-stoichiometric composition of Ni-Mn-Sb Heusler systems have been drawing much interest due to the presence of magneto-structural transition and many interesting magnetic properties associated with this transition, which occurs near room temperature.[8,9] With the aim of finding out novel and potential materials from this series, we have doped nonmagnetic elements Si and Ga in Sb site in $Ni_{46}Co_4Mn_{38}Sb_{12}$ alloy, resulting in the series $Ni_{46}Co_4Mn_{38}Sb_{12-x}Z_x$ (Z = Si and Ga ). The results of this study are presented in this paper.

## II. EXPERIMENTAL DETAILS

Polycrystalline $Ni_{46}Co_4Mn_{38}Sb_{12-x}Z_x$ (Z = Si and Ga ) ingots were prepared by arc melting process under argon atmosphere using Ni, Co, Mn, Sb and Si or Ga of atleast 99.99% purity. About 3 % extra manganese was added to compensate the weight loss and the final weight loss was found to be less than 1%. For better homogeneity the sample was melted four times and consequently annealed in evacuated quartz tube at 850°C for 24 hours. The structural characterization was done by powder x-ray diffraction (XRD) using Cu-Kα radiation. The magnetization measurements were carried out using a vibrating sample magnetometer attached to a Physical Property Measurement System (Quantum Design, PPMS-6500).

## III. RESULTS AND DISCUSSIONS

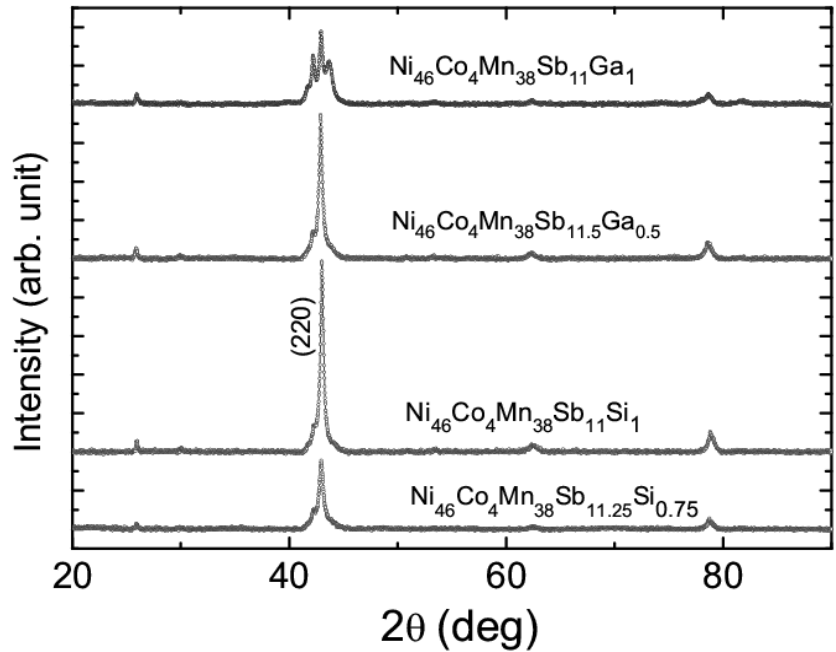

FIG.1. X-ray diffraction patterns of $Ni_{46}Co_4Mn_{38}Sb_{12-x}$ (Si, Ga)$_x$ alloys.

From X-ray diffraction (XRD) patterns of $Ni_{46}Co_4Mn_{38}Sb_{12-x}Si_x$ alloys (Fig. 1), it is found that alloys with x=0.75 and 1 show predominantly austenite (cubic) phase at room temperature. In $Ni_{46}Co_4Mn_{38}Sb_{12-x}Ga_x$ alloys, the one with x=0.5 shows predominantly austenite (cubic) phase, whereas, the one with x=1 shows predominantly martensite (orthorhombic) phase at room temperature. However, the existence of small peak near the (220) L2$_1$ peak in Ni$_{46}$Co$_4$Mn$_{38}$Sb$_{10.25}$Si$_{0.75}$ and Ni$_{46}$Co$_4$Mn$_{38}$Sb$_{11.5}$Ga$_5$ indicates the presence of a very small amount of martensite phase, as this phase starts just below the room temperature. In case of Ni$_{46}$Co$_4$Mn$_{38}$Sb$_{11}$Ga$_1$ the signature of (220) peak indicates the presence of small amount of austenite phase, as the transition exists just above the room temperature. The complete structural detail with martensitic transition in similar type of alloys is reported in Ref. No. 10.

FIG. 2. Temperature variation of magnetization in ZFC and FCC modes in a field of 1 kOe for (a) x=0.75, (b) x=1 in Ni$_{46}$Co$_4$Mn$_{38}$Sb$_{12-x}$Si$_x$ alloys and for (c) x=0.5, (d) x=1 in Ni$_{46}$Co$_4$Mn$_{38}$Sb$_{12-x}$Ga$_x$ alloys.

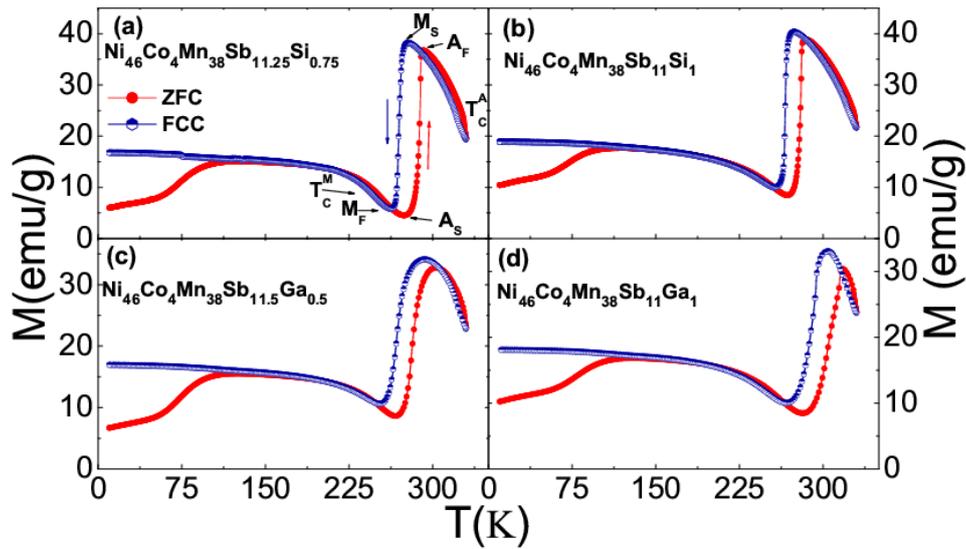

FIG. 2. Temperature variation of magnetization in ZFC and FCC modes in a field of 1 kOe for (a) x=0.75, (b) x=1 in Ni$_{46}$Co$_4$Mn$_{38}$Sb$_{12-x}$Si$_x$ alloys and for (c) x=0.5, (d) x=1 in Ni$_{46}$Co$_4$Mn$_{38}$Sb$_{12-x}$Ga$_x$ alloys.

Figure 2 shows magnetization versus temperature [M(T)] plots of Ni$_{46}$Co$_4$Mn$_{38}$Sb$_{12-x}$Z$_x$ (Z=Si and Ga) alloys. The magnetization was measured in zero field cooled (ZFC) and field cooled cooling (FCC) modes with 1kOe applied field. In the ZFC mode, the sample was cooled from 330 K to 10 K in the absence of magnetic field. Subsequently, a field of 1kOe was applied and the measurement was done by increasing temperature up to 330 K. In the FCC mode, without removing the field the measurement was performed on decreasing the temperature. As temperature decreases below 330 K, the magnetization reaches its maximum at $M_S$ (martensitic start temperature). By further decreasing the temperature, the austenite phase loses its stability and the magnetization suddenly decreases to $M_F$ (martensitic finish temperature). In this transition region there is mixed austenite and martensite phase present in the system. Below $M_F$, there exists a magnetic order-disorder transition, called the Curie temperature of the martensite phase $T_C^M$. Similarly, in the heating curve the temperature corresponding to the lowest magnetization is called $A_S$ (austenite start temperature), after which the martensite phase loses its stability and suddenly jumps to a maximum value at $A_F$ (austenite finish temperature). The ferro to paramagnetic transition occurs at the Curie temperature of the austenite phase $T_C^A$. Hysteresis between cooling and heating curves near the martensitic transition indicates the first order nature of transition. On the other hand, the separation between ZFC and FCC curves indicates the presence of AFM exchange interaction at low temperatures. The AFM component arises due to the off- stoichiometric composition of Ni-Mn-Sb.[11]

It can be observed from figure 2(a) and (b) that with Si substitution the $M_S$ and $M_F$ temperatures decrease. In figure 2(a), $M_F$ occurs at 258 K for Si=0.75, which decreases to 254 K with Si=1 in $Ni_{46}Co_4Mn_{38}Sb_{12-x}Si_x$ alloys. However, with Ga substitution the $M_S$ and $M_F$ values are larger. This suggests that with Si substitution the austenite phase gets stabilized, while with Ga the austenite phase loses its stability. These observations are consistent with XRD data discussed earlier. It can also be seen from Fig 2 that the martensitic transition is relatively sharper in the case of Si than that in Ga.

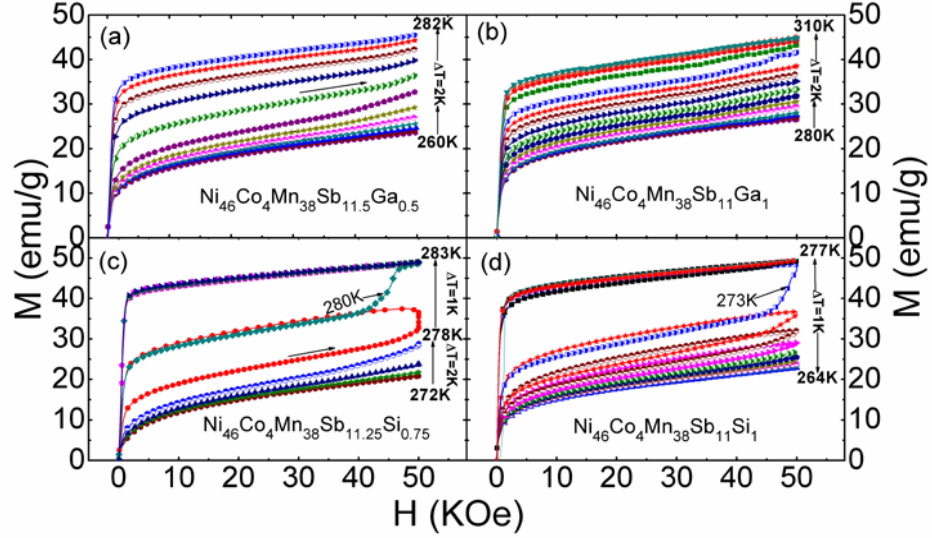

FIG. 3. Field dependence of magnetization for (a) $Si_{0.75}$, (b) $Si_1$, (c) $Ga_{0.5}$ and (d) $Ga_1$ in $Ni_{46}Co_4Mn_{38}Sb_{12-x}Z_x$ (Z=Si and Ga) alloys.

Isothermal *M(H)* curves for $Ni_{46}Co_4Mn_{38}Sb_{12-x}Z_x$ (Z=Si and Ga) alloys are shown in Fig. 3. For each alloy the *M(H)* was taken from 0 to 50 kOe in their respective structural transition region. In Si doped alloys, data were taken in both increasing and decreasing fields. For the sake of clarity these data are shown for certain temperatures only. It is observed that the magnetization increases with increase in temperature. This is because low temperature martensite phase possesses lower magnetization than the high temperature austenite phase. It is also observed that the magnetic isotherms measured above the martensitic transition temperature follow more or less the same magnetization path. For Si substitution with x=0.75, metamagnetic transition is observed at 278 K, 279 K and 280 K. For x=1 this is observed in 270 – 274 K temperature regime. This metamagnetic behaviour is generally observed close to $A_S$ and it vanishes at higher temperatures. However, for Ga substitution, less pronounced metamagnetic behaviour can be noticed in Fig. 3(c) and (d). For x=0.5 metamagnetic transition is observed at 272 K and 274 K. For x=1 this is observed near 302 K. The sharper metamagnetic transition in Si doping can be explained from the sharp martensitic transition in the system which is observed in *M(T)* plot. The non saturation of the magnetization curve reflects the presence of AFM components in the FM matrix.

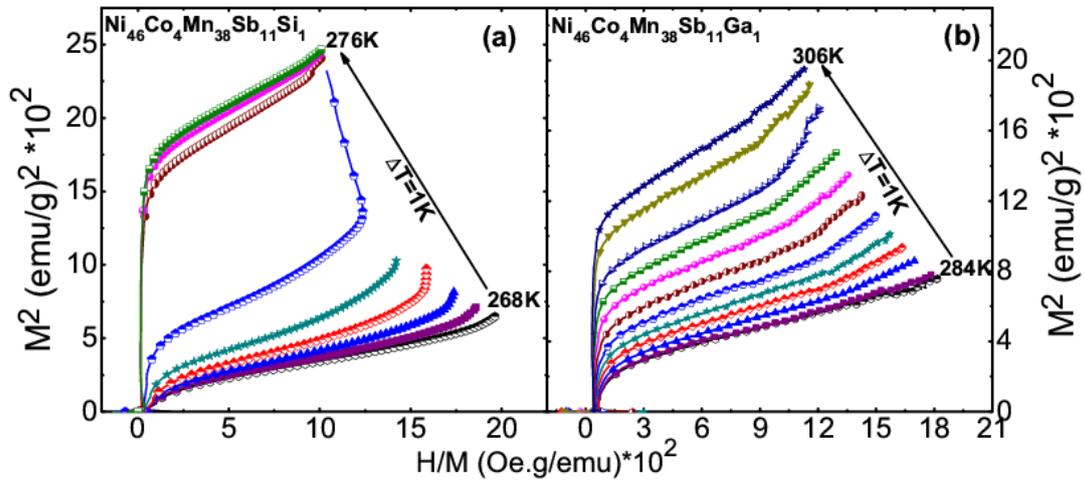

FIG. 4. Arrott plot for (a) $Ni_{46}Co_4Mn_{38}Sb_{11}Si_1$ and (b) $Ni_{46}Co_4Mn_{38}Sb_{11}Ga_1$.

The Arrott plot ($M^2$ vs. H/M) is shown in Fig. 4, which is calculated from M-H isotherms. The plots are shown for temperatures around the magneto-structural transition. It can be seen that the $M^2$ vs. H/M curves are S shaped. Generally S shaped Arrott plots indicate the first order nature of the transition.[12] The sharp metamagnetic transitions seen in the Si doped alloys is reflected in the S shaped Arrott plots shown in Fig. 4(a). However, for Ga doped sample (Fig. 4(b)) the Arrott plot is not S shaped, which implies that the transition is not of first order type in this case. Arrott plot have been employed to investigate the order of transition in NiMnGa alloys earlier as well.[13]

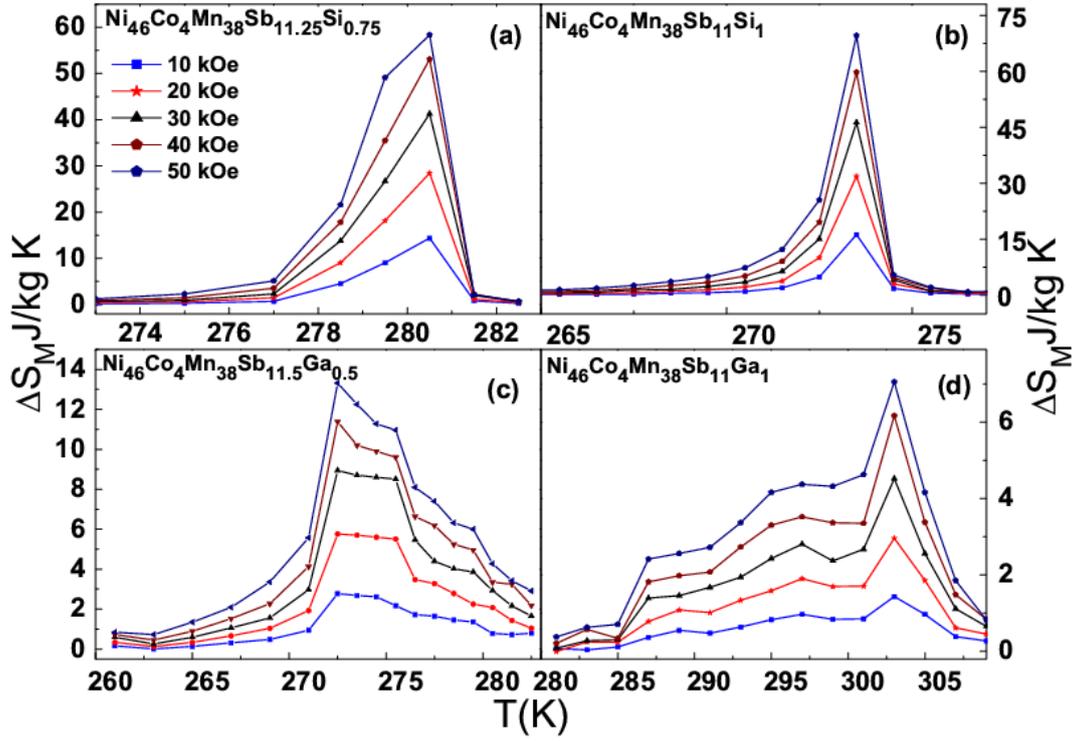

FIG. 5. Temperature dependence of the magnetic entropy change under different magnetic fields for (a) $Ni_{46}Co_4Mn_{38}Sb_{11.25}Si_{0.75}$, (b) $Ni_{46}Co_4Mn_{38}Sb_{11}Si_1$, (c) $Ni_{46}Co_4Mn_{38}Sb_{11.5}Ga_{0.5}$ and (d) $Ni_{46}Co_4Mn_{38}Sb_{11}Ga_1$ alloys.

Figure 5 shows inverse magnetocaloric effect, in terms of isothermal magnetic entropy change versus temperature calculated in the vicinity of the martensitic transition. Magnetic entropy change $\Delta S_M$ can be defined as

$$\Delta S_M(T, H) = S_M(T, 0) - S_M(T, H) \tag{1}$$

The isothermal entropy variation obtained using the Maxwell relation:[14,15]

$$\left(\frac{\partial S(T,H)}{\partial H}\right)_T = \left(\frac{\partial M(T,H)}{\partial T}\right)_H \tag{2}$$

After integrating equation (2), the equation can be written as

$$\Delta S_M(T,H) = \int_0^H \left(\frac{\partial M(T,H)}{\partial T}\right)_H dH \tag{3}$$

In Fig. 5(a) and (b) for *x*=0.75 and 1, maximum positive magnetic entropy change of 58 J/kg K and 70 J/kg K are observed at 280.5 K and 273.5 K respectively for field change of 50 kOe in $Ni_{46}Co_4Mn_{38}Sb_{12-x}Si_x$ alloys. It can be noted here that in $Ni_{46}Co_4Mn_{38}Sb_{12}$, $(\Delta S_M)_{max}$ of 43 J/kg K is reported.[16] Therefore, with addition of Si the maximum MCE value increases considerably, which is due to the increase in the *(∂M/∂T)* value at the martensitic transition. Strong magnetostructural coupling brought about by Si is the reason for this increase. However with Ga substitution, the MCE behaviour is different and the magnetic entropy change is quite low compared to that of $Ni_{46}Co_4Mn_{38}Sb_{12-x}Si_x$. For x=0.5 and 1, $\Delta S_M$ of 13 J/kg K and 7 J/kg K are observed at 272.5 K and 303 K respectively with Ga substitution. However, for Ga substituted alloys the MCE peak much broader compared to the Si doped alloys. The differences in the MCE data between Si and Ga substituted alloys clearly reflect the differences in the magnetization behaviour between them.

## IV. CONCLUSION

The effect of Si and Ga substitution in Sb site for $Ni_{46}Co_4Mn_{38}Sb_{12-x}Z_x$ (Z=Si and Ga) has been discussed by various magnetization measurements. From *M(T)* plot it is found that with Si substitution austenite phase get stabilized, however, with Ga substitution austenite phase loose its stability. Considerably large metamagnetic behaviour is observed in *M(H)* curve with Si substitution as compared to Ga substitution. Large magnetization difference is observed between the austenite and martensite phases with Si substitution. Enhanced MCE in the case of Si substituted alloys reflects the presence of strong magnetostructural coupling in the system. The Si substituted alloys show large and narrow MCE peaks, whereas, less pronounced and broad peaks are observed in the Ga substituted alloys. Thus, these alloys are found to be promising for fundamental studies and as well as for magnetic refrigeration applications.